\documentclass[preprint,showpacs,preprintnumbers,amsmath,amssymb]{revtex4}
\usepackage{graphicx,epsfig,dcolumn,bm,epic,eepic,float}
\usepackage{amsmath}
\usepackage{latexsym}
\usepackage{booktabs}
\usepackage{makeidx,shortvrb,latexsym}
\usepackage{color}
\begin{document}
\title{Application of the constituent quark exchange model to the parton distributions and the EMC ratios of $^{12}C$ and $^{14}N$ nuclei}
\author{A. Hadian }\altaffiliation {ahmad.hadian@ut.ac.ir}
\author{M. Modarres }\altaffiliation {mmodares@ut.ac.ir}
\affiliation{Physics Department, University of  Tehran, 1439955961,
Tehran, Iran.}
\begin{abstract}
The quark exchange model (QEM) is reformulated for the $A=12$ and $14$ systems to obtain the constituent quark distributions of $^{12}C$ and $^{14}N$ nuclei, respectively. Afterwards, the different types of the point-like parton distribution functions (PDFs), i.e., the valence quarks, the sea quarks and the gluons, are extracted from the constituent quark model (CQM), at the hadronic scale $Q_0^2=0.34$ $GeV^2$ to generate the constituent quark exchange model (CQEM = QEM $\oplus$ CQM) PDFs. From the resulted PDFs, the structure functions and the European Muon Collaboration (EMC) ratios of the above-mentioned nuclei at the appropriate hadronic scale are calculated. To make our results more comparable with the available experimental data, we evolve the PDFs, by using the standard Dokshitzer–Gribov–Lipatov–Altarelli–Parisi (DGALP) evolution equations, to the higher hard scales. Subsequently, the EMC ratios of $^{12}C$ and $^{14}N$ nuclei at the some higher energy scales, corresponding to the scales of available data, are calculated, at the leading (LO) and the next-to-leading (NLO) orders of pQCD. By doing so, it is observed that the EMC ratios do not significantly depend on the hard scale $Q^2$, and the outcomes are consistent with the various experimental data such as HERMES, BCDMS, JLab, SLAC, NMC, and EMC. Especially, in the mid-range of Bjorken $x$ values, the results are so desirable. Like our previous works for the $^4He$ and $^6Li$ nuclei, we again  observe that at a fixed
scale $Q^2$, the LO and the NLO EMC ratios with high precision are approximately the same. Therefore, one can conclude that at a given hard scale, the LO approximation is good enough for calculating the EMC ratios of light nuclei.
\end{abstract}
\pacs{ 13.60.Hb, 21.45.+v, 14.20.Dh, 24.85.+P, 12.39.Ki\\ Keywords:
Quark exchange model, Parton distribution functions, Constituent quark model, 
Structure function, EMC ratio, DGLAP evolution equations.} \maketitle
\section{Introduction}
The detailed analysis of leptons deep-inelastic scattering off nuclei reveals an appearance of an important  nuclear effect that rules out the naive picture of a nucleus as being a system of quasi-free nucleons. This effect was first announced by the European Muon
Collaboration (EMC) group, i.e., Aubert et al. \cite{Aubert}, in 1983. After this discovery, several approaches based on the traditional nuclear properties were suggested to justify the differences between the free and bound nucleon structure functions (SFs) \cite{Taylor,Sargsian,Piller,Frankfurt,Lamp,
Frankfurt.,Weinstein.,Melnitchouk.,Feynman,Close,Roberts,Afnan,Petratos,Bissey,Guzey,Malace}. Among these, the Fermi motion and the nuclear binding, the role of $\Delta$ particles, the quark exchange effect and the shadowing phenomena are most appealing to theoretical nuclear physicists. Very recently, via the KMR scheme in the "$k_t$-factorization" framework \cite{Kimber, Mod}, we investigated the contributions of "un-integrated" parton distribution functions (UPDFs) on the EMC ratios of $^3H$, $^3He$, $^4He$ and $^6Li$ nuclei \cite{Hadian5, Hadian4}, and it was demonstrated that the inclusion of these UPDFs can explain the reduction of EMC ratio at the small $x$ values, i.e., the shadowing region.

While the Fermi motion and the shadowing phenomena were proposed for the large and small Bjorken $x$ regions, respectively, the roles of binding, the $\Delta$ particles and the quark exchange model (QEM) were proposed for the mid-range $x$ values. The later was first established by Hoodbhoy and Jaffe for the three nucleon system to justify the EMC ratio of the hypothetical $^3He$-$^3H$ nucleus \cite{Jaffe,Hoodbhoy}. This formalism was recently expanded by us, with inclusion of the constituent quark model (CQM), into the four and six-nucleons systems to calculate the SFs and the EMC ratios of $^4He$ \cite{Hadian3} and $^6Li$ nuclei \cite{Hadian1,Hadian2}, respectively. In the following section, it will be reformulated for the heavier nuclear systems, such as $^{12}C$ and $^{14}N$ nuclei.

The quark exchange model (QEM) becomes very complicated if one includes the sea quark and gluon degrees of freedom. Therefore, one needs an additional approach to overcome this shortcoming. The constituent quark model (CQM) which was initially introduced by Feynman \cite{Feynman,Close,Roberts}, is a straightforward solution. Based on the CQM, the "constituent quarks" are supposed to be the complex
objects which consist of the point-like quarks, the anti-quarks, and the gluons as their components. We briefly review this model in the section III.

Therefore, in the present study, we first introduce the QEM for the $A=12$ and $14$ systems to obtain the quark momentum distributions of $^{12}C$ and $^{14}N$ nuclei, respectively (see the section II). Afterwards, to obtain the various point-like PDFs of these nuclei, the concept of CQM is presented in the section III. Subsequently, the SF and the EMC ratio calculations of the $^{12}C$ and $^{14}N$ nuclei in the frame-work of constituent quark exchange model (CQEM = QEM $\oplus$ CQM), at the leading (LO) and the next-to-leading (NLO) orders of pQCD, are given in the section IV. Eventually, the section V is devoted to the results, discussion, and conclusions.
\section{The QEM for the $\mathcal{A}=12$ and $14$ nuclear systems}
In this section, we intend to introduce the QEM for the $^{12}C$ and $^{14}N$ nuclei. One must note that all approximations that were made in the references \cite{Jaffe,Hoodbhoy}, are also considered here. Especially, the Fermi motion is excluded from the following calculation by regarding the lowest order expansion of the nuclear wave function. Therefore, as we expect, the final EMC results are not valid at large $x$ values, i.e., $x$ $\geq$ 0.8. In addition, we drop terms in which the simultaneous quark exchanges occur among more than the two correlated nucleons. 

We begin with defining a single nucleon state which composed of three quarks, as follows \cite{Betz,Jaffe,Owns,Hoodbhoy,Modarres,GHAFOORI}:
\begin{equation}\label{1}
\left| \alpha\right>={\mathcal{N}^{\alpha^\dagger}}\left|0\right>={1\over\sqrt{3!}}
\mathcal{N}^\alpha_{\mu_1\mu_2\mu_3}q^\dagger_{\mu_1}q^\dagger_{\mu_2}q^\dagger_{\mu_3}\left|0\right>.
\end{equation}
$\alpha$ = $\{\vec{p},M_S,M_T\}$, $\vec{p}$, $M_S$, and $M_T$ are the momentum, the spin and isospin projections of the nucleons, respectively. Similarly, $\mu_i$ = $\{\vec{k},m_s,m_t,c\}$, $\vec{k}, m_s, m_t$, and $c$ denote the momentum, the spin and isospin projections, and the color of quarks, respectively. $q_\mu^\dagger$ ($q_\mu$) and ${\mathcal{N}^{\alpha^\dagger}}$ (${\mathcal{N}^{\alpha}}$) are the creation (annihilation) operators, and a repeated index means a summation over all values of the coordinates as well as integration over momenta. $\mathcal{N}^\alpha_{\mu_1\mu_2\mu_3}$ is the totally anti-symmetric nucleon wave function, which is defined as:
\begin{equation}\label{2}
\mathcal{N}^\alpha_{\mu_1\mu_2\mu_3} =
D(\mu_1,\mu_2,\mu_3;\alpha_i){\times}\delta(\vec{k_1}+\vec{k_2}+\vec{k_3}-
\vec{P})\varphi(\vec{k_1},\vec{k_2},\vec{k_3},\vec{P}),
\end{equation}
where the nucleon wave function $\varphi(\vec{k_1},\vec{k_2},\vec{k_3},\vec{P})$ is approximated by a Gaussian:
\begin{equation}\label{3}
{\varphi(\vec{k_1},\vec{k_2},\vec{k_3},\vec{P})}=\Big({3b^4\over\pi^2}\Big)^{3\over4}exp{\Big[{-b^2(k_1^2+k_2^2+k_3^2)\over2}}+{{b^2P^2}\over6}\Big],
\end{equation}
in which $b$ is the nucleon radius.   $D(\mu_1,\mu_2,\mu_3;\alpha_i)$ (in the equation (\ref{2})) is the product of four Clebsch-Gordon coefficients $C^{{j_1}{j_2}{j}}_{{m_1}{m_2}{m}}$:
\begin{align}\label{4}
D(\mu_1,\mu_2,\mu_3;\alpha_i)=&{1\over\sqrt{3!}}\epsilon_{{c_1}{c_2}{c_3}}{1\over\sqrt{2}}\sum\limits_{s,t=0,1}
C^{{1\over2}{s}{1\over2}}_{{m_{s_\sigma}}{m_s}{M_{S_{\alpha_i}}}}C^{{1\over2}{1\over2}{s}}_{{m_{s_\mu}}{{m_{s_\nu}}}{m_s}}
C^{{1\over2}{t}{1\over2}}_{{m_{t_\sigma}}{m_t}{M_{T_{\alpha_i}}}}C^{{1\over2}{1\over2}{t}}_{{m_{t_\mu}}{{m_{t_\nu}}}{m_t}},
\end{align}
where $\epsilon_{{c_1}{c_2}{c_3}}$ is the color factor. Finally, the nucleus states are written as:
\begin{equation}\label{5}
\left|\mathcal{A}_i=n \right>
={1\over\sqrt{n!}}\chi^{{\alpha_1}{\alpha_2}{\alpha_3}\cdots{\alpha_{n-1}}{\alpha_{n}}}\:\mathcal{N}^{\alpha_1^\dagger}\mathcal{N}^{\alpha_2^\dagger}\mathcal{N}^{\alpha_{3}^\dagger}\cdots\mathcal{N}^{\alpha_{n-1}^\dagger}\mathcal{N}^{\alpha_{n}^\dagger}\left|
0 \right>.
\end{equation} 
$n$ is the nucleons number, i.e., $n$ = 12 (14) for $^{12}C$ ($^{14}N$) nucleus, and $\chi^{{\alpha_1}{\alpha_2}{\alpha_3}\cdots{\alpha_{n-1}}{\alpha_{n}}}$ is the nuclear wave function. To avoid more complexities, we can suppose both $^{12}C$ and $^{14}N$ nuclei as uniform systems and perform all calculations at the nuclear matter density. Therefore, the nuclear wave function can be chosen from the reference \cite{GHAFOORI}. However, it has been shown that the EMC results are not dependent on the choice of nuclear wave function, considerably \cite{Afnan,Bissey,Chen.,Stadler}. 

Now, we can calculate the constituent quark momentum distribution in the $A=n$ system via the following equation:
\begin{equation}\label{6}
\rho_{m_t}^{M_T}(\vec{k},\mathcal{A}_i)={{\left< \mathcal{A}_i=n
\right| q^\dagger_{\bar{\mu}}q_{\bar{\mu}} \left| \mathcal{A}_i=n \right>}\over
{\left< \mathcal{A}_i=n | \mathcal{A}_i=n \right>}}.
\end{equation}
The sign bar means no summation on $M_T$, $m_t$, and integration over $\vec{k}$
on the repeated index $\mu$. The above equation leads to similar mathematical expressions for $^{12}C$ and $^{14}N$ nuclei except that the coefficients of the results are different (because of the difference between the nucleon number of $^{12}C$ and $^{14}N$ nuclei). Therefore, in what follows, we denote the coefficients with $a_i$ ($i$ = 1, 2, ...,10) parameters. The values of $a_i$ are introduced later on (see the table 1). So, the denominator of equation (\ref{6}) can be written as follows:
\begin{equation}\label{7}
{\left< \mathcal{A}_i=n | \mathcal{A}_i=n
\right>}=\chi^{{\star}{\alpha_1}{\alpha_2}{\alpha_3}\cdots{\alpha_{n-1}}{\alpha_{n}}}\:\chi^{{\alpha_1}{\alpha_2}{\alpha_3}\cdots{\alpha_{n-1}}{\alpha_{n}}}
-{\left< \mathcal{A} | \mathcal{A} \right>}_{exch},
\end{equation}
where 
\begin{align}\label{8}
{\left< \mathcal{A} | \mathcal{A}
\right>}_{exch}=a_1\:\chi^{{\star}{\alpha_1}{\alpha_2}{\alpha_3}\cdots{\alpha_{n-1}}{\alpha_{n}}}
\:&\mathcal{N}^{\alpha_2}_{\mu_1\mu_2\mu_3}\mathcal{N}^{\beta_2}_{\mu_2\mu_3\rho_1}
\mathcal{N}^{\alpha_3}_{\rho_1\rho_2\rho_3}
\mathcal{N}^{\beta_3}_{\mu_1\rho_2\rho_3}\nonumber\\&\times\delta^{{\alpha_1}{\beta_1}}\delta^{{\alpha_4}{\beta_4}}\cdots\delta^{{\alpha_{n-1}}{\beta_{n-1}}}\delta^{{\alpha_{n}}{\beta_{n}}}
\:\chi^{{\beta_1}{\beta_2}{\beta_3}\cdots{\beta_{n-1}}{\beta_{n}}}.
\end{align}
After performing some more algebraic operations, the nominator of equation (\ref{6}) takes the following form:
\begin{align}\label{9}
&{\left< \mathcal{A}_i=n \right| q^\dagger_{\mu}q_{\mu} \left|
\mathcal{A}_i=n \right>}
=a_2\:\chi^{{\star}{\alpha_1}{\alpha_2}\cdots{\alpha_{n-1}}{\alpha_{n}}}
\Big(\mathcal{N}^{\alpha_1}_{\mu\sigma_2\sigma_3}\mathcal{N}^{\beta_1}_{\mu\sigma_2\sigma_3}
\delta^{{\alpha_2}{\beta_2}}\delta^{{\alpha_3}{\beta_3}}\cdots\delta^{{\alpha_{n-1}}{\beta_{n-1}}}\delta^{{\alpha_{n}}{\beta_{n}}}\nonumber\\&\quad-\big(a_3\:\mathcal{N}^{\alpha_1}_{\mu\sigma_2\sigma_3}\mathcal{N}^{\beta_1}_{\mu\sigma_2\sigma_3}
\mathcal{N}^{\alpha_2}_{\mu_1\mu_2\mu_3}\mathcal{N}^{\beta_2}_{\rho_1\mu_2\mu_3}
\mathcal{N}^{\alpha_3}_{\rho_1\rho_2\rho_3}\mathcal{N}^{\beta_3}_{\mu_1\rho_2\rho_3}+
a_4\:\mathcal{N}^{\alpha_2}_{\mu\mu_1\mu_2}\mathcal{N}^{\beta_2}_{\mu\mu_2\rho_1}\mathcal{N}^{\alpha_3}_{\rho_1\rho_2\rho_3}
\mathcal{N}^{\beta_3}_{\mu_1\rho_2\rho_3}\nonumber\\&\quad\quad\quad\quad\quad\quad+
a_5\:\mathcal{N}^{\alpha_2}_{\mu_1\mu_2\mu_3}\mathcal{N}^{\beta_2}_{\mu\mu_2\mu_3}\mathcal{N}^{\alpha_3}_{\mu\rho_2\rho_3}
\mathcal{N}^{\beta_3}_{\mu_1\rho_2\rho_3}\big)
\delta^{{\alpha_4}{\beta_4}}\cdots\delta^{{\alpha_{n-1}}{\beta_{n-1}}}\delta^{{\alpha_{n}}{\beta_{n}}}\Big)
\chi^{{\beta_1}{\beta_2}\cdots{\beta_{n-1}}{\beta_{n}}}.
\end{align}
Finally, after performing all summations over the repeated indices of equation (\ref{6}), we get the iso-scalar quark momentum distribution (averaged on $\mathcal{M_T}$):
\begin{equation}\label{10}
\rho(k)={{\rho_{dir}(k)+\rho_{exch}(k)}\over
{\Big[1+{a_6}\:\mathcal{I}\Big]}},
\end{equation}
where
$$\rho_{dir}(k)={a_7}\:A(k),$$
$$\rho_{exch}(k)={a_8}\:B(k)+{a_9}\:C(k)+{a_{10}}\:D(k).$$
The direct coefficient $A(k)$ and exchange terms $B(k)$, $C(k)$, $D(k)$ and the overlap integral $\mathcal{I}$ are given in the equations $(4)-(8)$ of the reference \cite{Hadian5}, respectively.
The values of coefficients $a_i$ for $^{12}C$ and $^{14}N$ nuclei, are illustrated in the table 1.
From the above equations, as required by the quark number conservation for a nucleus with mass number $\mathcal{A}=n$ = 12 or 14, it can easily be shown that:
\begin{equation}\label{11}
\int{\rho(k)d{\vec{k}}}={n\over2},
\end{equation}
which confirms the analytic calculations and the consistency of the approximations we have made for the $\mathcal{A}=n$ iso-scalar system. 
\section{The PDFs of $^{12}C$ and $^{14}N$ nuclei in the CQM}
In this section, we initially present the concept of CQM. Based on CQM, the constituent quarks are complex objects
and their SFs are described by a set of functions, $\phi_{q_1q_2}(x)$, that traditionally are called the SFs of the constituent quarks. $\phi_{q_1q_2}(x)$ identifies the number of point-like partons of type $q_2$, which are present in the constituent of type $q_1$ with fraction $x$ of its total momentum.

First, we aim to briefly explain how the SFs of the constituent quarks can be attained. The figure 1 (diagram $(a)$) illustrates the general representation of the SF of a constituent as the $s$-channel absorptive part of a forward current-constituent scattering amplitude. In the gluon model, such an absorptive part can be separated into three classes as shown in the  figure 1 (the diagrams $(b)-(d)$). The diagrams of the first class contribute only if the parton $q_2$ is of the same type as the constituent $q_1$, while in the second class of diagrams $q_2$ = $\bar{q}_1$. The  diagrams of type $(d)$ give contributions that are independent both from $q_2$ and $q_1$. Although both diagrams of type $(b)$ and $(c)$ in the figure 1 contain a quark-antiquark pair in the t-channel, the second type contributions are much smaller due to the duality requirement. Therefore, the contributions of type $(c)$ will be neglected. In conclusion, correspond to the diagrams of type $(b)$ and $(d)$ we can consider the following equations, respectively \cite{Altarelli}:
\begin{equation}\label{12}
\phi^{(b)}_{q_1q_2}(x)=\delta_{q_1q_2}\:\phi^{(b)}(x),
\end{equation}
\begin{equation}\label{13}
\phi^{(d)}_{q_1q_2}(x)=\phi^{(d)}(x),
\end{equation}
in which:
\begin{equation}\label{14}
\phi^{(b)}(x)\underset{x\rightarrow 0}{\sim} {B\over \sqrt{x}},
\end{equation}
\begin{equation}\label{15}
\phi^{(d)}(x)\underset{x\rightarrow 0}{\sim} {C\over{x}}.
\end{equation}
The contributions $(b)$ and $(d)$ simply correspond to the leading parton (of the same type of the constituent itself) and the sea pairs 
distribution, respectively. A similar behavior is also expected to hold for the probability distribution of the neutral gluons:
\begin{equation}\label{16}
\phi_{qg}(x)\underset{x\rightarrow 0}{\sim}{G\over{x}}.
\end{equation}
Based on the equations (\ref{15}) and (\ref{16}), the average number of soft sea partons of any kind inside a given constituent can be assumed as follow:
\begin{eqnarray}\label{17}
&dN(x)\underset{x\rightarrow 0}{\sim}A\:{dx\over{x}}, \nonumber\\&A=G+6C. 
\end{eqnarray}
From the above equations, the $x\rightarrow 1$ behavior of the leading parton distribution, $\phi^{(b)}_{q_1q_2}(x)$, can be derived from the fundamental assumption that: the probability of finding the leading parton with a
fraction $x$ of the constituent momentum is equivalent to the probability that all the other partons carry away exactly a fraction $(1- x)$. Therefore, the general functional form of $\phi^{(b)}(x)$ at the large $x$ region can be written as follow: 
\begin{equation}\label{18}
\phi^{(b)}(x)\underset{x\rightarrow 1}{\sim} {(1-x)}^{A-1}.
\end{equation}
Additionally, $\phi^{(b)}(x)$ must satisfy the following normalization condition:
\begin{equation}\label{19}
\int{\phi^{(b)}(x)}\:dx=1.
\end{equation}
From the equations (\ref{14}), (\ref{18}) and (\ref{19}) it can be concluded that:
\begin{equation}\label{200}
\phi^{(b)}(x)={{\Gamma(A+{1\over2})}\over{\Gamma({1\over2})\Gamma(A)}}{{({1-{x})}^{A-1}} \over \sqrt{x}}.
\end{equation}
On the other hand, the $x\rightarrow 1$ behavior of $\phi^{(d)}_{q_1q_2}(x)$ and $\phi_{qg}(x)$ should not dominate over $\phi^{(b)}_{q_1q_2}(x)$. Consequently, the simple functional form for these probability distributions can be suggested as follows:
\begin{equation}\label{201}
\phi^{(d)}(x)={C\over{x}}\:{{({1-{x})}^{D-1}}},
\end{equation}
\begin{equation}\label{202}
\phi_{qg}(x)={G\over{x}}\:{{({1-{x})}^{B-1}}}.
\end{equation}
The average momentum carried by the second moments of the parton distributions is known experimentally. For example, at the hadronic scale $\mu_0^2 = 0.34\:GeV^2$, 53.5\% of the nucleon momentum is carried by the valence quarks, 35.7\% by the gluons and the remaining momentum is belonged to the sea quarks. This information is used to extract the value of the constant parameters in the above equations, i.e., $A$, $B$, $G$ and the ratio $C/D$. at the hadronic scale, it can be concluded that: 
$$A=0.435,\: B=0.378,\: C=0.05,\: D=2.778, \:G=0.135.$$
More information and detailed discussion about the above SFs for different kinds of partons, and the procedure of evaluating these constants can be found in the references \cite{Altarelli,Manohar,Scopetta..,Traini}.

Ultimately, at the hadronic scale $\mu_0^2$ = 0.34 $GeV^2$, the point-like parton distributions can be obtained from the constituent quark distributions via the following equation \cite{Altarelli,Manohar,Scopetta..}:
\begin{equation}\label{20}
q(x,\mu_0^2)=\int^{1}_{x}{dz\over
z}\Big[{\mathcal{U}(z,\mu_0^2)\phi_{\mathcal{U}q}\Big({x\over z},
\mu_0^2\Big)+\mathcal{D}(z,\mu_0^2)\phi_{\mathcal{D}q}\Big({x\over
z}, \mu_0^2\Big)}\Big],
\end{equation}
where $q$ labels the various point-like partons, i.e., the valence quarks ($u_v$, $d_v$), the sea
quarks ($u_s$, $d_s$, $s$), the sea anti-quarks ($\bar{u}_s$, $\bar{d}_s$, $\bar{s}$), and the gluons ($g$). Additionally, $\mathcal{U}$ and $\mathcal{D}$ specify the constituent
probability distributions of $u$ and $d$ quarks, respectively. In this approach, the sea quark and the anti-quark
distributions are independent of the iso-spin flavor. Therefore, in what
follows, the sea distributions are demonstrated by $q_s$. It should be noted that in the equation (\ref{20}), $\phi_{\mathcal{U}d}\Big({x\over z},
\mu_0^2\Big)$ and $\phi_{\mathcal{D}u}\Big({x\over z}, \mu_0^2\Big)$
are zero, because in the constituent quark of type $\mathcal{U}$,
there is no point-like valence quark of type $d$ and vice versa (see diagram $b$ of figure 1). 

According to the equation (\ref{20}), first, we need to calculate the constituent quark distribution functions. At each $Q^2$ scale, the constituent quark distributions in the nucleons of the nucleus $\mathcal{A}_i$, $f^j_a(x,Q^2;\mathcal{A}_i)$, can be related to the quark momentum distributions, $\rho^j_a(\vec{k};\mathcal{A}_i)$, as follows ($j$ = $p$,
$n$ ($a$ = $u$, $d$) for the proton (up quark) and the neutron (down quark), respectively) \cite{Rasti1,Rasti2,Gonzalez}:
\begin{equation}\label{21}
f^j_a(x,Q^2;\mathcal{A}_i)={{1}\over{(1-x)^2}}\int{\rho^j_a(\vec{k};\mathcal{A}_i)}\:\delta\Big({x\over{1-x}}-{k_{+} \over M}\Big)d\vec{k},
\end{equation}
where the constituent quark light-cone momentum in the rest frame of target is used with ${{k^0={({{\vec{k}^2}+m_a^2}})}^{1\over2}-{\epsilon}_0}$. The two free parameters, i.e., ${m_a}$ and ${\epsilon}_0$, are the quark masses and their binding energies, respectively. We can determine these free
parameters such that the best fit to the valence up quark distribution of Martin et al., i.e., MMHT 2014 \cite{Harland}, is achieved. However, as it was shown in the most of the previous related works, the final EMC results are not sensitive to these parameters, considerably   \cite{Betz,Jaffe,Owns,Hoodbhoy,Modarres,GHAFOORI,Modares,Zolfagharpour.,Yazdanpanah.,
Rasti,Rujula,Ebrahimi,Yazdan,Yazdan.,Yazdan..}. The constituent distributions are finally obtained by performing the angular integration in the equation (\ref{21}) as follows:
\begin{equation}\label{22}
f^j_a(x,Q^2;\mathcal{A}_i)={{2\pi
M_T}\over{(1-x)^2}}\int^{\infty}_{k^a_{min}}{{{\rho^j_a(k;\mathcal{A}_i)}}kdk},
\end{equation}
with,
\begin{equation}\label{23}
k^a_{min}(x)={({{x M_T\over{1-x}}+{\epsilon^a_0})^2}-m^2_a\over{2({x
M_T\over{1-x}}+{\epsilon^a_0})}},
\end{equation}
where $M_T$ denotes the nucleon mass. Therefore, the up and down constituent quark distributions can be obtained via the following equations, respectively:
\begin{equation}\label{24}
\mathcal{U}(z,\mu_0^2;\mathcal{A}_i)={{2\pi
M_T}\over{(1-z)^2}}\int^{\infty}_{k^u_{min}}{{{\rho_{\mathcal{U}}(k;\mathcal{A}_i)}}kdk},
\end{equation}
\begin{equation}\label{25}
\mathcal{D}(z,\mu_0^2;\mathcal{A}_i)={{2\pi
M_T}\over{(1-z)^2}}\int^{\infty}_{k^d_{min}}{{{\rho_{\mathcal{D}}(k;\mathcal{A}_i)}}kdk}.
\end{equation}
However, for each iso-scalar nucleus, the up and down quark momentum
distributions are equal, i.e., $\rho_{\mathcal{U}}(k)$ = $\rho_{\mathcal{D}}(k)$ = $\rho(k)$. Accordingly, the up and down constituent quark distributions of an iso-scalar system are equal and can be obtained as follows:
\begin{equation}\label{26}
\mathcal{U}(z,\mu_0^2;\mathcal{A}_i)=\mathcal{D}(z,\mu_0^2;\mathcal{A}_i)={{2\pi
M_\mathcal{N}}\over{(1-z)^2}}\int^{\infty}_{k_{min}}{{{\rho(k;\mathcal{A}_i)}}kdk},
\end{equation}
where we use the averaged mass for the proton and neutron, i.e.,
$M_T$ = $M_\mathcal{N}$ = $(M_P + M_N)\over{2}$. By substituting the quark momentum
distributions of $^{12}C$ and $^{14}N$ nuclei from equation (\ref{10}) in the above equation, we can calculate their constituent quark distributions, respectively. 

In summary, the various PDFs of $^{12}C$ and $^{14}N$ nuclei at the hadronic scale $\mu_0^2$ can be obtained as follows (for $\mathcal {A}$ = 12 and 14, the superscript $^{\mathcal {A}}X$ denotes $^{12}C$ and $^{14}N$, respectively):
\begin{equation}\label{27}
u_v^{^{\mathcal {A}}X}(x,\mu_0^2)=d_v^{^{\mathcal {A}}X}(x,\mu_0^2)=\int^{1}_{x}{dz\over
z}{\mathcal{U}^{^{\mathcal {A}}X}(z,\mu_0^2)\phi_{\mathcal{U}q_v}\Big({x\over z},
\mu_0^2\Big)},
\end{equation}
\begin{equation}\label{28}
q_s^{^{\mathcal {A}}X}(x,\mu_0^2)=2\int^{1}_{x}{dz\over
z}{\mathcal{U}^{^{\mathcal {A}}X}(z,\mu_0^2)\phi_{\mathcal{U}q_s}\Big({x\over z},
\mu_0^2\Big)},
\end{equation}
\begin{equation}\label{29}
g^{^{\mathcal {A}}X}(x,\mu_0^2)=2\int^{1}_{x}{dz\over
z}{\mathcal{U}^{^{\mathcal {A}}X}(z,\mu_0^2)\phi_{\mathcal{U}g}\Big({x\over z},
\mu_0^2\Big)}.
\end{equation}
These resulted PDFs of $^{12}C$ and $^{14}N$ nuclei, for the nucleon's radius $b$ = 0.8 $fm$ and at the hadronic scale $\mu_0^2$ = 0.34 $GeV^2$, are displayed in the panels (a) and (b) of the figure 2, respectively. $(m_a,\:\epsilon_0^a)$ are chosen as (325, 140 $MeV$) for the $^{12}C$ and (330, 140 $MeV$) for the $^{14}N$. By determination of the various PDFs of $^{12}C$ and $^{14}N$ nuclei, we are now able to calculate the NLO (LO) SFs and NLO (LO) EMC ratios of these iso-scalar light nuclei, which is the main discussion of the following section.
\section{The SF and EMC ratio calculations}
In this section, we aim to present the formulation of the SF and EMC ratio at the NLO limit for $^{12}C$ and $^{14}N$ nuclei. In the NLO limit, the specific target SF, $\mathcal{F}_2^{\mathcal{A}_i}(x,Q^2)$, in terms of the 
point-like valence quark, sea quark, and gluon distributions can be obtained via the following equation \cite{Gluck,Nemat}:
\begin{eqnarray}\label{30}
&&\mathcal{F}_2^{\mathcal{A}_i}(x,Q^2)=x\sum\limits_{a=u,d,s\:;\:j=p,n}Q\:_a^{2}\:\Big[\:q\:^a_j(x,Q^2)+\bar{q}\:^a_j(x,Q^2)+\:{{\alpha_s(Q^2)}\over{2\pi}}\int_x^1{dy\over{y}}\nonumber\\&&{\quad}{\quad}{\quad}{\quad}{\quad}{\quad}\times\big[\:C_{q,2}({x\over y})\:\big(q\:^a_j(y,Q^2)+\bar{q}\:^a_j(y,Q^2)\big)+2\:C_{g,2}({x\over y})\:g(y,Q^2)\big]\Big],
\end{eqnarray}
with
\begin{equation}\label{300}
C_{q,2}(z)={4\over 3}\Big[{1+z^2\over {1-z}}\Big(\ln{{1-z\over {z}}-{3\over 4}}\Big)+{1\over 4}(9+5z)\Big]_+,
\end{equation}
\begin{equation}\label{301}
C_{g,2}(z)={1\over 2}\Big[(z^2+(1-z)^2)\:\ln{1-z\over {z}}-1+8z(1-z)\Big].
\end{equation}
Note that $\alpha_s(Q^2)$, the NLO coupling constant, is defined as follows:
\begin{equation}\label{31}
{{\alpha_s}\over{4\pi}}\cong{1\over{{\beta_0}\log({Q^2\over\Lambda^2})}}-{{{\beta_1}\ln\ln({Q^2\over\Lambda^2})}\over{{\beta_0^3}\:{[\ln({Q^2\over\Lambda^2})]}^2}},
\end{equation}
where $\beta_0$ and $\beta_1$ are the first two universal coefficients of the QCD $\beta$ functions, 
and $\Lambda$ is taken to be 0.248 $GeV$ throughout our calculations \cite{Gluck}. The LO expressions are obviously entailed in the above equations by simply dropping all higher order terms $(C_{i,2},\beta_i)$ in (\ref{30}) and (\ref{31}) \cite{Gluck}. According to the equation (\ref{30}), the NLO SFs of $^{12}C$ and $^{14}N$ nuclei can be written via the following equation (as we mentioned, for $\mathcal {A}$ = 12 and  14, the superscript $^{\mathcal {A}}X$ denotes $^{12}C$ and $^{14}N$, respectively):
\begin{eqnarray}\label{32}
\mathcal{F}_2^{^{\mathcal {A}}X}(x,Q^2)=&&x\Big[{a^{\prime}_1}u_v^{^\mathcal {A}X}(x,Q^2)+{{a^{\prime}_2}}\:d_v^{^\mathcal {A}X}(x,Q^2)+{{a^{\prime}_3}}\:q_s^{^\mathcal {A}X}(x,Q^2)\nonumber\\&&+{{\alpha_s(Q^2)}\over{2\pi}}\int_x^1{dy\over{y}}\Big(C_{q,2}({x\over y})\:\big({{a^{\prime}_4}}u_v^{^\mathcal {A}X}(y,Q^2)+{{a^{\prime}_5}}\:d_v^{^\mathcal {A}X}(y,Q^2)+{{{a^{\prime}_6}}}\:q_s^{^\mathcal {A}X}(y,Q^2)\big)\nonumber\\&&+{{{a^{\prime}_7}}}\:C_{g,2}({x\over y})\:g^{^\mathcal {A}X}(y,Q^2)\Big)\Big],
\end{eqnarray}
where $u_v^{^\mathcal {A}X}$, $d_v^{^\mathcal {A}X}$, $q_s^{^\mathcal {A}X}$, and $g^{^\mathcal {A}X}$ denote the valence up, down and sea quarks, and the gluon distributions of $^\mathcal {A}X$ (= ${^{12}C}$ or ${^{14}N}$) nucleus, respectively. The numerical values of $a^{\prime}_i$ ($i$ = 1,2,...,7) coefficients for $^{12}C$ and $^{14}N$ nuclei are given in the table 2. As it was mentioned before, for each iso-scalar nucleus the up and down valence quark distributions are equal. As a consequent, the NLO SFs of $^{12}C$ and $^{14}N$ nuclei can be finally written as follows:
\begin{eqnarray}\label{33}
&&\mathcal{F}_2^{^{\mathcal {A}}X}(x,Q^2)=x\Big[{({a^{\prime}_1+a^{\prime}_2})}\:u_v^{^{\mathcal {A}}X}(x,Q^2)+{{a^{\prime}_3}}\:q_s^{^{\mathcal {A}}X}(x,Q^2)+{{\alpha_s(Q^2)}\over{2\pi}}\int_x^1{dy\over{y}}\nonumber\\&&\times\Big(C_{q,2}({x\over y})\big[({a^{\prime}_4+a^{\prime}_5})\:u_v^{^{\mathcal {A}}X}(y,Q^2)+{{a^{\prime}_6}}\:q_s^{^{\mathcal {A}}X}(y,Q^2)\big]+{{a^{\prime}_7}}\:C_{g,2}({x\over y})\:g^{^{\mathcal {A}}X}(y,Q^2)\Big)\Big],
\end{eqnarray}

The NLO (LO) EMC ratio, $\mathcal{R}_{EMC}$, is defined as the ratio of bound nucleon's NLO (LO) SF in the target (obviously per nucleon, such as ${^{12}C}$ or ${^{14}N}$) to that of the free nucleon. Therefore, $\mathcal{R}_{EMC}$ can be obtained as \cite{Jaffe}:
\begin{equation}\label{34}
\mathcal{R}_{EMC}={\mathcal{F}^{T}_2(x)\over{\mathcal{F}^{T^\star}_2(x)}}.
\end{equation}
$T$ indicates the target averaged over nuclear spin and isospin and $T^\star$ denotes to the hypothetical target with exactly the same quantum numbers, but in which their nucleons are sufficiently far away from each other such that any quark exchange does not occur between them. Note that the effects of nuclear Fermi motion are excluded from both $T$ and $T^\star$ and the deviation of $\mathcal{R}$ from unity is only due to the exchange of constituent quarks between the static nucleons. According to the equation (\ref{34}), we can calculate the EMC ratios of ${^{12}C}$ and ${^{14}N}$ nuclei, both at the NLO and LO levels. The results are presented in the next section.
\section{Results, Discussion and Conclusions}
The NLO (LO) EMC ratios of  ${^{12}C}$ nucleus at the energy scale $\mu_0^2$ = 0.34 $GeV^2$ and the different values of $b$ = 0.7, 0.8, and 0.9 $fm$ are illustrated in the figure 3 (the full curves). There are two main points about this figure. First, like our previous works \cite{Hadian2,Hadian3}, we again observe that for each value of nucleon's radius $b$, the resulting LO and NLO EMC ratios at $\mu_0^2$ = 0.34 $GeV^2$ are almost equal (since the curves that are obtained
from the LO and NLO EMC calculations overlap, one can consider the full
curves in the figure 3 as the LO or the NLO EMC ratios). This is justified that in the EMC calculations, in which we consider the ratio of bound and free nucleon's SFs, this ratio at the LO level is approximately equal to that of the NLO level. Second, as the values of $b$ increase from 0.7 $fm$ to 0.9 $fm$, the corresponding curves deviate more from unity. As it was mentioned in our previous reports \cite{Hadian1,Hadian2,Hadian3,Hadian4,Hadian5}, this obvious behavior is due to the increase in the probability of quark exchange among the constituent quarks, which considerably affects the deviation of EMC ratio. It should be noted that because of the omission of the Fermi motion effect, these curves monotonically decrease and the enhancement of the EMC ratios at large values of $x$ is not achieved. Therefore, in what follows, the EMC curves are plotted for $x$ $\leq$ 0.75. For comparison, the available experimental data \cite{Malace} are also displayed in the Figure 3. These measurements are from JLab \cite{Seely2} (the filled circles), NMC \cite{Arneodo} (the filled triangles), SLAC \cite{Gomez} (the filled squares), and EMC (the filled stars) \cite{Arneodo2}. It is seen that our theoretical results are significantly consistent with the various experimental data. However, in the small $x$ region, due to omission of both the shadowing phenomena and the transverse momentum effect, i.e., the UPDF contributions, our outcomes do not show complete agreement.

TO make our results more comparable with the experimental data, we must evaluate the EMC ratios of ${^{12}C}$ nucleus at some higher energy scales in which the experiments are conducted. To this end, the PDFs of ${^{12}C}$ nucleus (at $b$ = 0.8 $fm$) are evolved by using the solution of DGLAP equations \cite{Gribov,Lipatov,Altarelli1,Dokshitzer} given in the reference \cite{Botje}, i.e., the QCDNUM code. Afterwards, the LO and NLO EMC ratios of ${^{12}C}$ nucleus are computed for those Bjorken $x$ values and $Q^2$ energy scales, in which the experimental data exist. The NLO (LO) EMC results are shown in the figure 4 (the hollow squares, circles, triangles, and stars in the panels (a), (b), (c), and (d), respectively). In this figure, the hollow squares, circles,  triangles and stars represent the NLO (LO) EMC calculations (the NLO and LO EMC results overlap) for the different values of $x$ and $Q^2$, corresponding to those of SLAC \cite{Malace,Gomez} (the filled squares in the panel (a)), JLab \cite{Malace,Seely2} (the filled circles in the panel (b)), NMC \cite{Malace,Arneodo} (the filled triangles in the panel (c)), and EMC \cite{Malace,Arneodo2} (the filled stars in the panel (d)), respectively. it is observed that at a fixed scale $Q^2$, the LO and NLO EMC results are almost equal. The same conclusion was made in our previous works \cite{Hadian2,Hadian3}. In the panel (c) of figure 4, the EPPS16 fit results \cite{Eskola} (the filled rhombuses) with their upper and lower bands (the full curves) are also presented for comparison. A great agreement is observed between the theoretical and experimental results. In addition, as we concluded in our previous works \cite{Hadian4, Hadian5}, it is noticeable that the EMC ratio is not significantly dependent on the energy scale $Q^2$.

The figure 5 shows the NLO (LO) ratio of SF of ${^{12}C}$ nucleus (from the present study) to that of ${^{6}Li}$ nucleus (from the reference \cite{Hadian2}) at the hadronic scale $\mu_0^2$ = 0.34 $GeV^2$ and for the nucleon's radius $b$ = 0.8 $fm$ (the full curve). The filled squares are the corresponding NMC experimental data from the reference \cite{Armadruz}. Again, to make our results more consistent with the experimental data, the LO and NLO ratios are calculated for those $x$ and $Q^2$ values in which the NMC data exist (the hollow squares). The filled circles represent the EPPS16 fit and the dashed curves are the lower and upper bands of EPPS16 analysis \cite{Eskola}. An acceptable consistency between our results and the experimental data is observed. Furthermore, the ratio does not depend on the energy scale, considerably. As we concluded before, in this case the LO and NLO ratios are again approximately equal.

The figure 6 illustrates the NLO (LO) EMC ratios of  ${^{14}N}$ nucleus at the energy scale $\mu_0^2$ = 0.34 $GeV^2$ and for different values of $b$ = 0.7, 0.8, and 0.9 $fm$ (the full curves in the panel (a)). It is observed again that for each value of nucleon's radius $b$, the resulting LO and NLO EMC ratios at $\mu_0^2$ = 0.34 $GeV^2$ are almost the same (the full curves in this figure represent both the NLO and LO results). Furthermore, as one expects, the curves corresponding to the larger values of $b$ have greater deviations from unity, which is due to the increase in the probability of quark exchange among the constituent quarks. The experimental measurements shown in this figure are from HERMES \cite{Malace,Airapetisn} (the filled circles), and BCDMS \cite{Malace,Bari} (the filled squares). To provide more consistent results, we again evaluate the LO and NLO EMC ratios (at $b$ = 0.8 $fm$) for various $x$ and $Q^2$ values in which the experimental data exist. These calculations are presented via the hollow circles (see the panel (b)) and the hollow squares (see the panel (c)) corresponding to the HERMES \cite{Malace,Airapetisn} and BCDMS \cite{Malace,Bari} data, respectively. In all cases, our results are in agreement with data, properly. It can be again concluded that the LO and NLO ratios with high precision are the same (the NLO and LO EMC results overlap). Additionally, as discussed in the previous works  \cite{Hadian4,Hadian5,Gomez}, the EMC ratio again does not significantly depend on the energy scale.

The comparisons of EMC ratios of ${^{4}He}$ (the full curve), ${^{6}Li}$ (the dotted curve), ${^{12}C}$ (the dash curve), and ${^{14}N}$ (the dash-dotted curve) nuclei at the energy scale $\mu_0^2$ = 0.34 $GeV^2$ and nucleon's radius $b$ = 0.8 $fm$ are displayed in the figure 7. The ${^{4}He}$ and ${^{6}Li}$ EMC ratios are taken from the reference \cite{Hadian3} and \cite{Hadian2}, respectively. As expected, by increasing the number of nucleons in the nucleus, the probabilities of quark exchanges among the nucleons are increased, which make the $\mathcal{R}_{EMC}$ to have greater deviation from unity. The similar conclusions were also made in our previous reports \cite{Hadian3,Hadian5}.

In conclusion, the QEF was reformulated for the $A$ = 12 and 14 systems to extract the quark momentum distributions of ${^{12}C}$ and  ${^{14}N}$ nuclei, respectively. Afterwards, by using the CQM, the different point-like PDFs of the these nuclei at the hadronic scale 0.34 $GeV^2$ were calculated. These PDFs were then evolved by DGLAP evolution equations \cite{Gribov,Lipatov,Altarelli1,Dokshitzer} to some higher energy scales corresponding to the experimental results. By using these resulted PDFs, the EMC ratios of the above-mentioned nuclei at the LO and NLO levels were then evaluated at the different energy scales. It was concluded that in all cases, the LO and the NLO EMC ratios with high precision are the same. As we mentioned before \cite{Hadian2,Hadian3}, this can indicate that for calculating the EMC ratios of the finite light nuclei, the LO approximation is good enough.  It was observed that in all cases, our EMC results were consistent with the various experimental data. However, for the small $x$ domain ($x$ $\leq$ 0.1), because of omission of both the shadowing effect and the UPDFs contributions, the results were unfavorable. Similar to our previous works \cite{Hadian2,Hadian3}, to estimate exclusively the dynamical effects of QEM, we used the density of nuclear matter for the calculation of the exchange integral $\mathcal{I}$. The Fermi motion effect should be considered separately (e.g. see reference \cite{Rasti}). We ignored this effect by the approximation which was made to obtain the exchange coefficients $B(k)$, $C(k)$, $D(k)$ and the overlap integral $\mathcal{I}$, and plotted all the EMC ratios for the $x$ $\leq$ 0.75 domain. However, we hope to consider the Fermi motion effect as well as the UPDFs contributions in our future works to improve the present EMC results in the large and small $x$ regions, respectively.
\section*{Acknowledgements}
$MM$ would like to acknowledge the Research Council of University of
Tehran for the grants provided for him.

\begin{center}
\begin{tabular}{c @{\hspace{0.8cm}} c @{\hspace{0.65cm}} c @{\hspace{0.65cm}} c @{\hspace{0.65cm}} c @{\hspace{0.65cm}} c @{\hspace{0.65cm}} c @{\hspace{0.65cm}}c @{\hspace{0.65cm}} c @{\hspace{0.65cm}}c @{\hspace{0.65cm}} c}
\hline
{\footnotesize} & {\normalsize $a_1$} &{\normalsize $a_2$} & {\normalsize $a_3$} & {\normalsize $a_4$} &{\normalsize $a_5$ }&{\normalsize $a_6$}&{\normalsize $a_7$}&{\normalsize $a_8$}&{\normalsize $a_9$} & {\normalsize $a_{10}$}\\
\hline
$^{12}C$ & 594 & 36 & 495 & 66 & 33 & $99 / 4$ & 18 & 110 & $44 / 3$ & $22 / 3$ \\
$^{14}N$ & 819 & 42 & 702 & 78 & 39 & $273 / 8$ & 21 & 182 & $182 / 9$ & $91 / 9$ \\
\hline
\end{tabular}
\end{center}
\begin{table}[ht]
\caption{The numerical values of $a_i$ parameters which are used in the section II for the calculations of quark momentum distributions of $^{12}C$ and $^{14}N$ nuclei.}
\end{table}
\begin{center}
\begin{tabular}{c @{\hspace{0.85cm}} c @{\hspace{0.7cm}} c @{\hspace{0.7cm}} c @{\hspace{0.7cm}} c @{\hspace{0.7cm}} c @{\hspace{0.7cm}} c @{\hspace{0.7cm}}c @{\hspace{0.7cm}}}
\hline
{\footnotesize} & {\normalsize $a^\prime_1$} &{\normalsize $a^\prime_2$} & {\normalsize $a^\prime_3$} & {\normalsize $a^\prime_4$} &{\normalsize $a^\prime_5$ }&{\normalsize $a^\prime_6$}&{\normalsize $a^\prime_7$}\\
\hline
$^{12}C$ & $10 / 3$ & $10 / 3$ & 16 & $10 / 3$ & $10 / 3$ & 16 & 4   \\
$^{14}N$ & $35 / 9$ & $35 / 9$ & $56 / 3$ & $35 / 9$ & $35 / 9$ & $56 / 3$ & $14 / 3$ \\
\hline
\end{tabular}
\end{center}
\begin{table}[ht]
\caption{The numerical values of $a^{\prime}_i$ parameters which are used in the section IV for the calculations of SFs of $^{12}C$ and $^{14}N$ nuclei.}
\end{table}
\begin{figure}[ht]
\includegraphics[width=15cm, height=15cm]{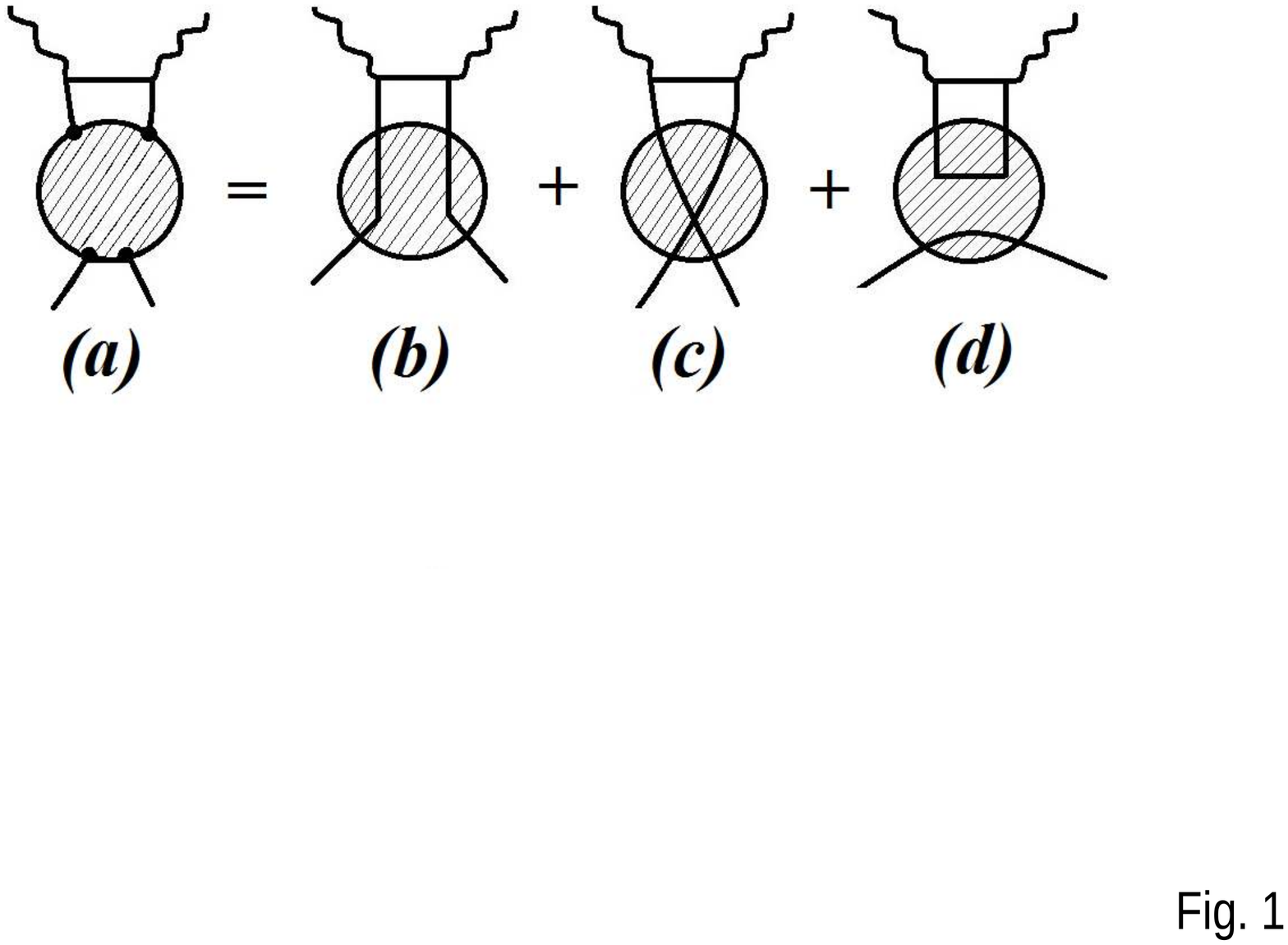}
\caption{The diagram $(a)$ is the general representation of SFs as the s-channel
absorptive part in the quark-parton model. The diagrams $(b)$, $(c)$ and $(d)$ show the different types of contributions to the SFs in the gluon-quark model.}
\end{figure}
\begin{figure}[ht]
\includegraphics[width=15cm, height=15cm]{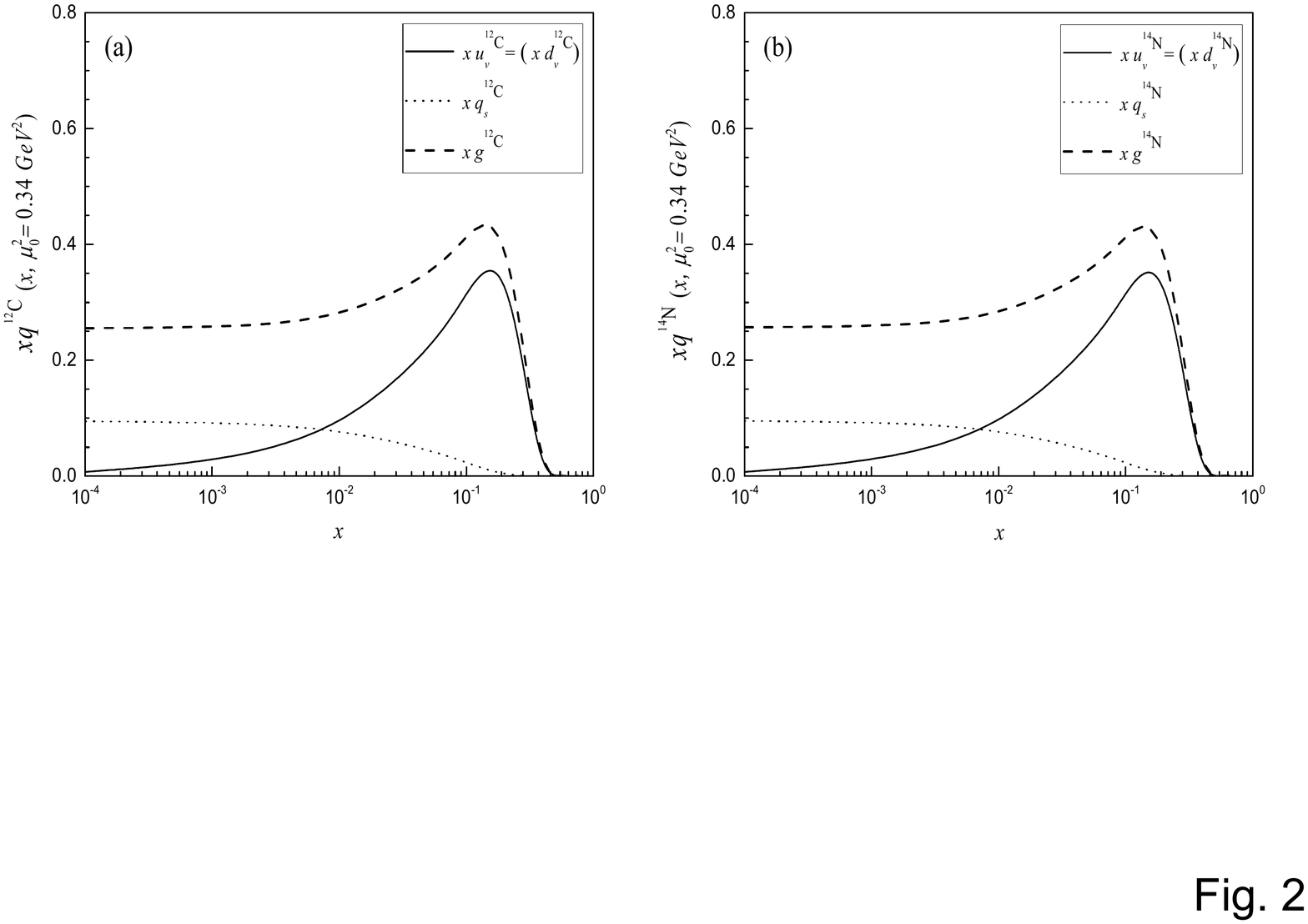}
\caption{The PDFs of $^{12}C$ (the panels (a)) and $^{14}N$ (the panels (b)) nuclei, for the nucleon's radius $b$ = 0.8 $fm$ and at the hadronic scale $\mu_0^2$ = 0.34 $GeV^2$. $(m_a,\:\epsilon_0^a)$ are chosen as (325, 140 $MeV$) for $^{12}C$ and (330, 140 $MeV$) for $^{14}N$ nuclei. In both panels, the dash curves represent the gluon distributions, while the solid and dotted
curves indicate the valence and sea quark distributions, respectively.}
\end{figure}
\begin{figure}[ht]
\includegraphics[width=15cm, height=15cm]{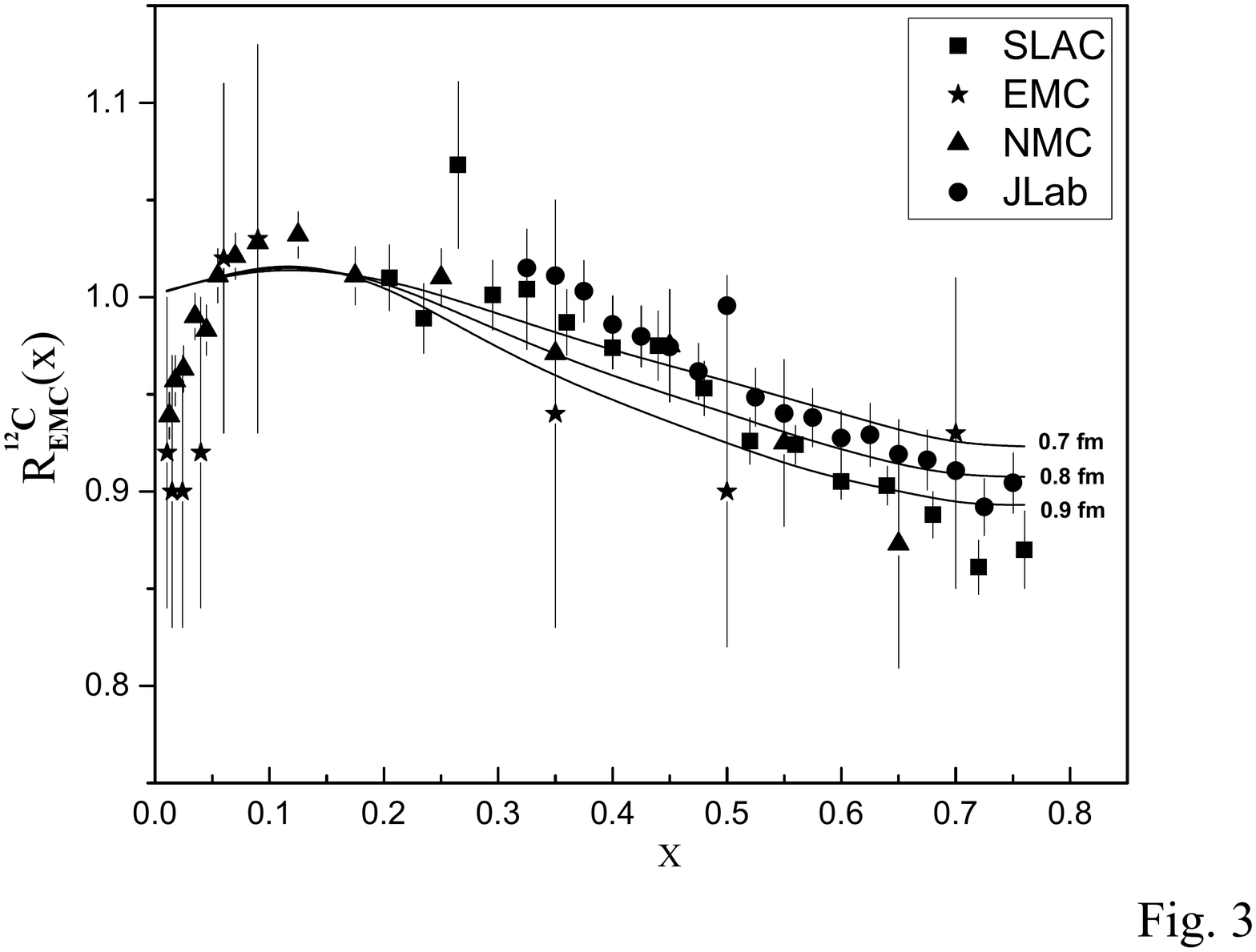}
\caption{The full curves are the NLO (LO) EMC ratio of $^{12}C$ nucleus for the different values of $b$ at $\mu_0^2$ = 0.34 $GeV^2$. The circles, the triangles, the squares, and the stars are from the JLab \cite{Malace,Seely2}, NMC \cite{Malace,Arneodo}, SLAC \cite{Malace,Gomez}, and EMC  \cite{Malace,Arneodo2} experimental data, respectively. Note that the experimental $Q^2$ values are different from 0.34 $GeV^2$ (see the figure 4).}
\end{figure}
\begin{figure}[ht]
\includegraphics[width=15cm, height=15cm]{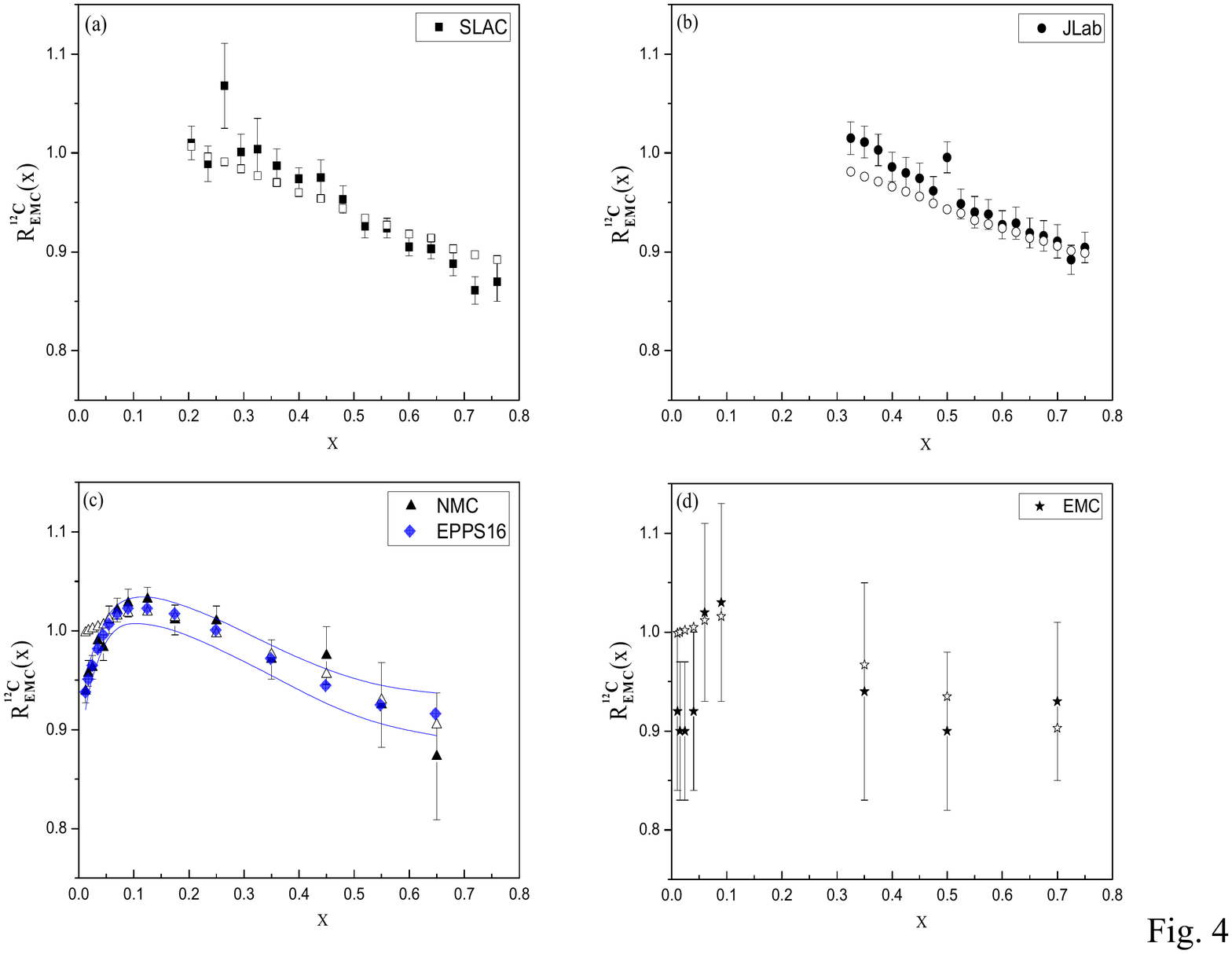}
\caption{The panels (a), (b), (c), and (d) are related to the comparison of theoretical EMC calculations of $^{12}C$ nucleus with those from the SLAC, JLab, NMC, and EMC, respectively. The filled squares in the panel (a), the filled circles in the panel (b), the filled triangles in the panel (c), and the filled stars in the panel (d) are from the SLAC \cite{Malace,Gomez}, JLab \cite{Malace,Seely2}, NMC \cite{Malace,Arneodo}, and EMC \cite{Malace,Arneodo2} experimental data, respectively. The hollow squares in the panel (a), the hollow
circles in the panel (b), the hollow triangles in the panel (c), and the hollow stars in the panel (d) are our NLO (LO) EMC calculations of $^{12}C$ nucleus at $b$ = 0.8 $fm$
for the different values of $x$ and $Q^2$, corresponding to those from the SLAC, JLab, NMC, and EMC data, respectively. The filled rhombuses and the full curves in the panel (c) are the EPPS16 fit results and their error bands \cite{Eskola}, respectively}
\end{figure}
\begin{figure}[ht]
\includegraphics[width=15cm, height=15cm]{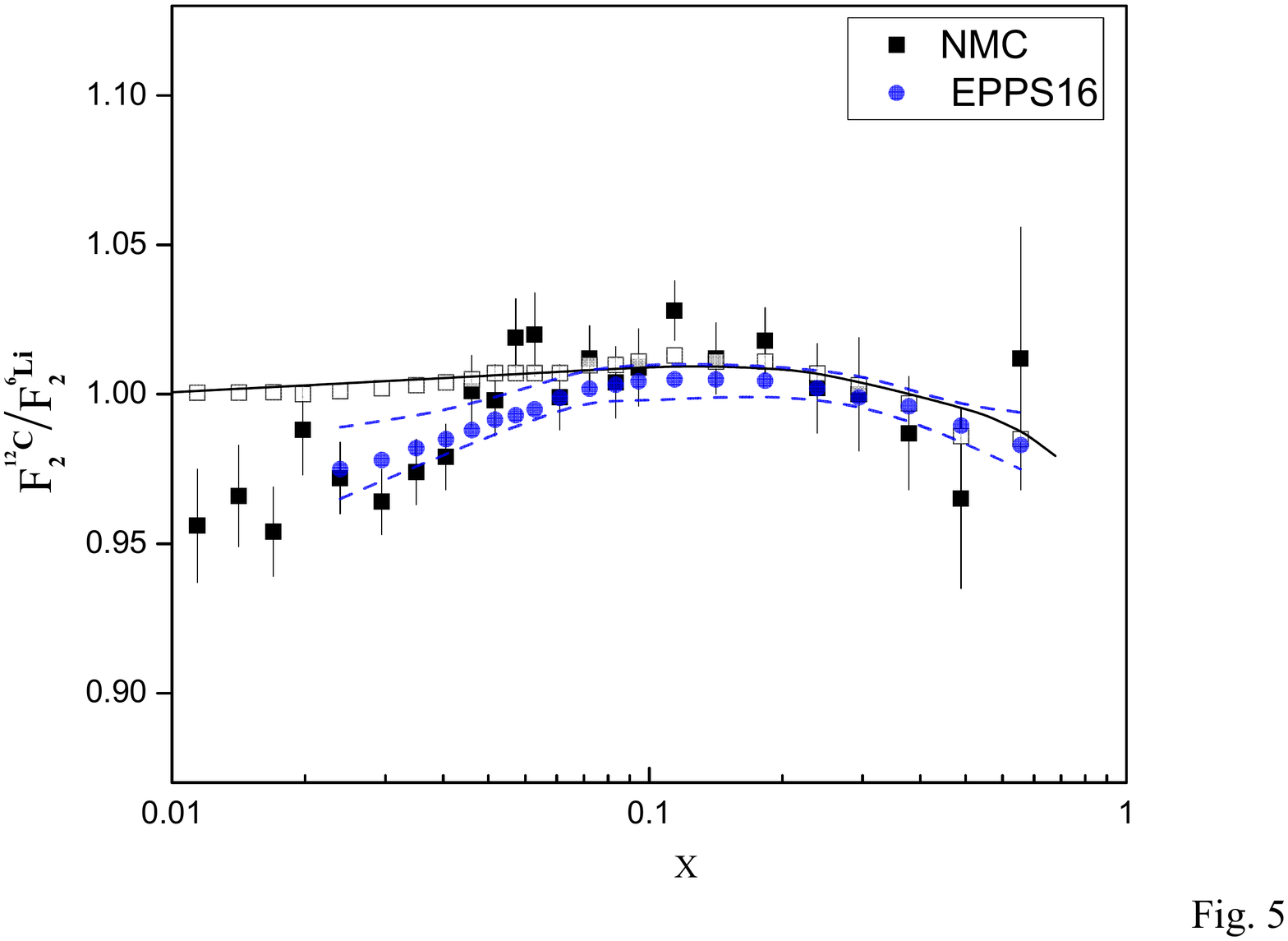}
\caption{The full curve represents the theoretical NLO (LO) ratio of SF of ${^{12}C}$ nucleus to that of ${^{6}Li}$ nucleus \cite{Hadian2} at the hadronic scale $\mu_0^2$ = 0.34 $GeV^2$ and nucleon's radius $b$ = 0.8 $fm$. The filled squares are from the NMC experimental data \cite{Armadruz}. The hollow squares are the theoretical NLO (LO) ratios  at $b$ = 0.8 $fm$
for the different values of $x$ and $Q^2$, corresponding to those from the NMC  experimental data. The filled circles and the dashed curves are the EPPS16 fit results and their error bands \cite{Eskola}, respectively}
\end{figure}
\begin{figure}[ht]
\includegraphics[width=15cm, height=15cm]{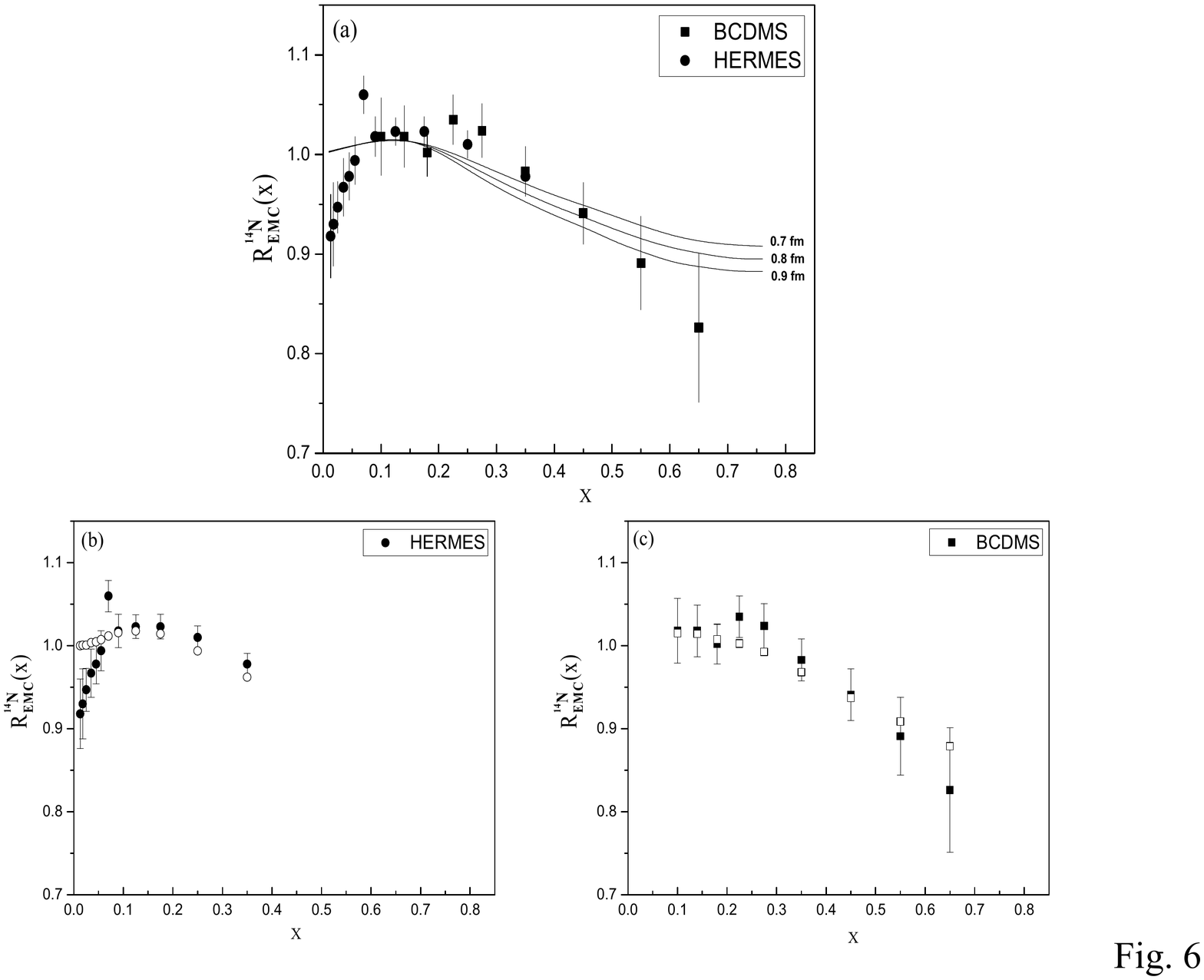}
\caption{In the panel (a), the full curves are the theoretical NLO (LO) EMC ratio of $^{14}N$ nucleus for the different values of $b$ and at a fixed energy scale, i.e., $\mu_0^2$ = 0.34 $GeV^2$. In this panel, the
circles, and the squares are from the HERMES \cite{Malace,Airapetisn} and BCDMS \cite{Malace,Bari} experimental data, respectively. The panels (b), and (c) are related to the comparison of theoretical  EMC calculations of $^{14}N$ nucleus (at the different energy scales) with those from the HERMES \cite{Malace,Airapetisn}, and BCDMS \cite{Malace,Bari} experimental data, respectively. The filled circles in the panel (b), and the filled squares in the panel (c) are from the HERMES \cite{Malace,Airapetisn}, and BCDMS \cite{Malace,Bari} collaborations, respectively. The hollow circles in the panel (b), and the hollow squares in the panel (c) are our NLO (LO) EMC calculations of $^{14}N$ nucleus at $b$ = 0.8 $fm$
for the different values of $x$ and $Q^2$, corresponding to those from the HERMES, and BCDMS data, respectively.}
\end{figure}
\begin{figure}[ht]
\includegraphics[width=15cm, height=15cm]{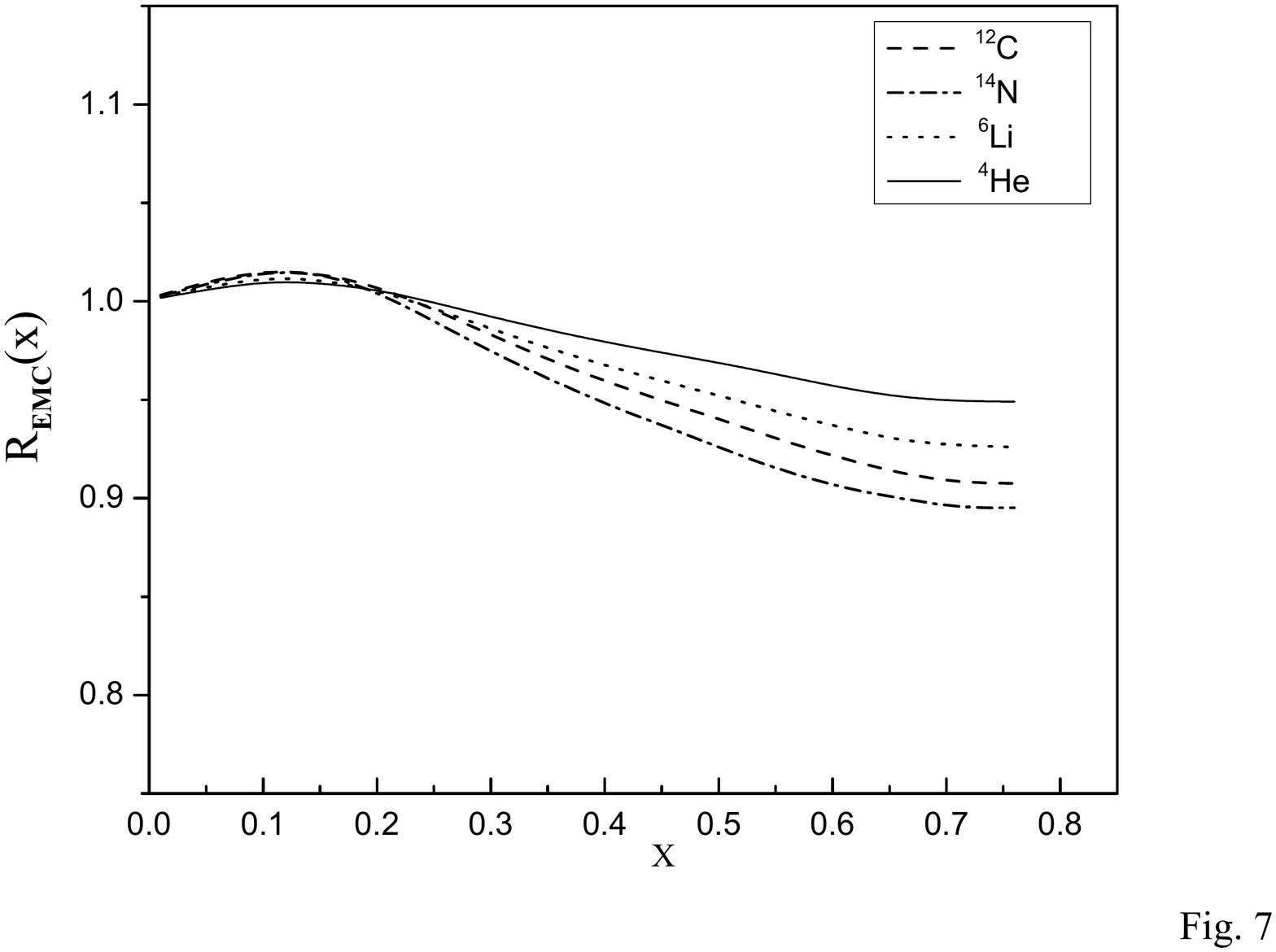}
\caption{The comparisons of EMC ratios of ${^{4}He}$ (the full curve), ${^{6}Li}$ (the dotted curve), ${^{12}C}$ (the dash curve), and ${^{14}N}$ (the dash-dotted curve) nuclei at the energy scale $\mu_0^2$ = 0.34 $GeV^2$ and the nucleon's radius $b$ = 0.8 $fm$. The ${^{4}He}$ and ${^{6}Li}$ EMC ratios are plotted from the reference \cite{Hadian3} and \cite{Hadian2}, respectively.}
\end{figure}

\begin{thebibliography}{99}
\bibitem{Aubert} J.J. Aubert et al., Phys. Lett. B 105 (1983) 403.
\bibitem{Taylor} R.E. Taylor, Rev. Mod. Phy. 63 (1991) 573.
\bibitem{Sargsian} M.M. Sargsian et al., J. Phys. G 29 (2003) R1.
\bibitem{Piller} G. Piller, W. Wesie, Phys. Rep. 330 (2000) 1.
\bibitem{Frankfurt} L.L. Frankfurt, M.I. Strikman, Phys. Rep. 76 (1981) 215.
\bibitem{Lamp} B. Lamp, E. Reya, Phys. Rep. 332 (2000) 1.
\bibitem{Frankfurt.} L. Frankfurt, M. Strikman, Int. J. Mod. Phys. E 21 (2012) 1230002.
\bibitem{Weinstein.} L.B. Weinstein et al., Phys. Rev. Lett 106 (2011) 052301.
\bibitem{Melnitchouk.} W. Melnitchouk, AIP Conf. Proc 1261 (2010) 85.
\bibitem{Feynman}R.P. Feynman, Benjamin, New York (1972).
\bibitem{Close} F.E. Close, Academic Press, London (1989).
\bibitem{Roberts} R.G. Roberts, Cambridge University Press, New York (1993).
\bibitem{Afnan} I.R. Afnan et al., Phys. Rev. C 68 (2003) 035201.
\bibitem{Petratos} G.G. Petratos et al., Temple University, Philadelphia, PA, USA (2000).
\bibitem{Bissey} F. Bissey, A.W. Thomas, I.R. Afnan, Phy. Rev. C 64 (2001) 024004.
\bibitem{Guzey} L. Frankfurt, V. Guzey, M. Strikman, Phys. Rep 512 (2012) 255.
\bibitem{Malace} S. Malace, D. Gaskell, D.W. Higinbotham, I.C. Cloet, Int. J. Mod. Phys. E 23 (2014) 1430013.
\bibitem{Kimber} M. A. Kimber, A. D. Martin, and M. G. Ryskin, Phys. Rev. D 63 (2001) 114027.
\bibitem{Mod} M. Modarres et al., J. Phys. G: Nucl. Part. Phys. 46 (2019) 105005.
\bibitem{Hadian5} M. Modarres, A. Hadian, Nuclear Physics A 983 (2019) 118.
\bibitem{Hadian4} M. Modarres, A. Hadian, Phys. Rev. D 98 (2018) 076001.
\bibitem{Jaffe} P. Hoodbhoy, R.L. Jaffe, Phys. Rev. D 35 (1987) 113.
\bibitem{Hoodbhoy} P. Hoodbhoy, Nucl. Phys. A 465 (1987) 113.
\bibitem{Hadian3} M. Modarres, A. Hadian, Eur. Phys. J. A 54 (2018) 236.
\bibitem{Hadian1} M. Modarres, A. Hadian, Int. J. Mod. Phys. E 24 (2015) 1550037.
\bibitem{Hadian2} M. Modarres, A. Hadian, Nucl. Phys. A 966 (2017) 342.
\bibitem{Betz} M. Betz, G. Krein, Th.A.J. Maris, Nucl. Phys. A 437 (1985) 509.
\bibitem{Owns} D.W. Duke, J.F. Owns, Phys. Rev. D 30 (1984) 49.
\bibitem{Modarres} M. Modarres, J. Phys. G: Nucl. Part. Phys. 20 (1994) 1423.
\bibitem{GHAFOORI} M. Modarres, K. Ghafoori-Tabrizi, J. Phys. G: Nucl. Phys. 14 (1988) 1479.
\bibitem{Chen.} C.R. Chen, G.L. Payne, J.L. Friar, B.F. Gibson, Phys. Rev. C 3 (1986) 1740.
\bibitem{Stadler} A. Stadler, W. Glockle, P.U. Sauer, Phys. Rev. C 44 (1991) 2319.
\bibitem{Altarelli} G. Altarelli, N. Cabibbo, L. Maiani, R. Petronzio, Nucl. Phys. B 69 (1974) 531
\bibitem{Manohar} A. Manohar, H. Georgi, Nucl. Phys. B 234 (1984) 189
\bibitem{Scopetta..} S. Scopetta, V. Vento, M. Traini, Phys. Lett. B 421 (1998) 64
\bibitem{Traini} M. Traini, V. Vento, A. Mair, and A. Zambarda, Nucl. Phys. A 614 (1997) 472.
\bibitem{Rasti1}  M. Modarres, M. Rasti, and M. M. Yazdanpanah, Few-Body Syst. 55 (2014) 85.
\bibitem{Rasti2}  M. M. Yazdanpanah, M. Modarres, and M. Rasti, Few-Body Syst. 48 (2010) 19.
\bibitem{Gonzalez} P. Gonzalez et al., Z. Phys. A 350 (1995) 371.
\bibitem{Harland} L.A. Harland-Lang, A.D. Martin, P. Motylinski, R.S. Thorne, 
Eur. Phys. J. C 75 (2015) 204.
\bibitem{Modares}M. M. Yazdanpanah and M. Modarres, Phys. Rev. C 57 (1998) 525.
\bibitem{Zolfagharpour.} M. Modarres and F. Zolfagharpour, Nucl. Phys. A 765 (2006) 112.
\bibitem{Yazdanpanah.} M. Modarres, M. Rasti, M. M. Yazdanpanah, Few-Body Syst. 55 (2014) 85.
\bibitem{Rasti} M. Modarres, M. Rasti, Int. J. Mod. Phys. E 22 (2013) 1350037.
\bibitem{Rujula} A. De Rujula, F. Martin, Phys. Rev. D 22 (1980) 1767.
\bibitem{Ebrahimi} E. Ebrahimi, M. Modarres, M.M. Yazdanpanah, Few-Body Syst. 39 (2006) 177.
\bibitem{Yazdan} M.M. Yazdanpanah, M. Modarres, Few-Body Syst. 37 (2005) 33.
\bibitem{Yazdan.} M.M. Yazdanpanah, M. Modarres, Eur. Phys. J. A 6 (1999) 91.
\bibitem{Yazdan..} M.M. Yazdanpanah, M. Modarres, Eur. Phys. J. A 7 (2000) 573.
\bibitem{Gluck} M. Gluck, E. Reya, A. Vogt, Z. Phys. C 67 (1995) 433.
\bibitem{Nemat} H. Nematollahi, M. M. Yazdanpanah, Phys. Rev. C 92 (2015) 015209.
\bibitem{Seely2} J. Seely, A. Daniel, D. Gaskell, J. Arrington et al., Phys. Rev. Lett. 103 (2009) 202301.
\bibitem{Arneodo} New Muon Collaboration. Collaboration (M. Arneodo et al.), Nucl. Phys. B 441 (1995) 12. 
\bibitem{Gomez}J. Gomez, R. Arnold, P. E. Bosted, C. Chang et al., Phys. Rev. D 49 (1994) 4348.
\bibitem{Arneodo2} European Muon Collaboration Collaboration (M. Arneodo et al.), Nucl. Phys. B 333 (1990) 1.
\bibitem {Gribov}V. N. Gribov and L. N. Lipatov, Yad. Fiz. 15 (1972) 781.
\bibitem{Lipatov} L. N. Lipatov, Sov. J. Nucl. Phys. 20 (1975) 94.
\bibitem{Altarelli1} G. Altarelli and G. Parisi, Nucl. Phys. B 126 (1977) 298.
\bibitem{Dokshitzer} Y. L. Dokshitzer, Sov. Phys. JETP 46 (1977) 641.
\bibitem{Botje} M. Botje, Comput. Phys. Commun. 182, 490 (2011), arXiv:1005.1481.
\bibitem{Eskola} K. J. Eskola1, P. Paakkinen, H. Paukkunen, C. A. Salgado, Eur. Phys. J. C 77 (2017) 163.
\bibitem{Armadruz} Armadruz et al. Nucl. Phys. B 441 (1995) 3.
\bibitem{Airapetisn} A. Airapetisn et al., Phys. Lett. B 567 (2003) 339.
\bibitem{Bari} BCDMS Collaboration (G. Bari et al.), Phys. Lett. B 163 (1985) 282.
\end{thebibliography}
\end{document}